\documentstyle[12pt,epsfig]{cernart} 
\makeatletter
\def\thebibliography#1{\section*{\refname\@mkboth
  {\uppercase{\refname}}{\uppercase{\refname}}}\list
  {\@biblabel{\arabic{enumi}}}{\settowidth\labelwidth{\@biblabel{#1}}%
  \leftmargin\labelwidth
  \advance\leftmargin\labelsep
  \usecounter{enumi}
  \def\theenumi{\arabic{enumi}}}%
  \def\newblock{\hskip .11em plus.33em minus.07em}%
  \sloppy\clubpenalty4000\widowpenalty4000
  \sfcode`\.=1000\relax}
\makeatother
\def\Aref#1{$^{\rm #1}$} 
\def\AAref#1#2{$^{\rm #1,#2}$}      
\def\IAref#1#2{$^{\Inst{#1},\rm #2}$}  
\def\IIref#1#2{$^{\Inst{#1},\Inst{#2}}$}

\def\Iref#1{$^{\Inst{#1}}$}  
\begin {document}
\begin {titlepage}
\docnum {CERN--PPE/95--31}
\date {September 29, 1995}

\hspace{11cm}hep-ex/9509014

\vspace{1cm}

\title {LARGE ENHANCEMENT OF DEUTERON POLARIZATION WITH FREQUENCY MODULATED
MICROWAVES}

\vspace{3.5cm}

\collaboration {The Spin Muon Collaboration (SMC)}


\vspace {3.5cm}

\begin{abstract}

We report a large enhancement of 1.7 in deuteron polarization up to values of
$0.6$ due to frequency modulation of the polarizing microwaves in a two liters
polarized target using the method of dynamic nuclear polarization. This target
was used during a deep inelastic polarized muon-deuteron scattering experiment
at CERN. Measurements of the electron paramagnetic resonance absorption spectra
show that frequency modulation gives rise to additional microwave absorption in
the spectral wings. Although these results are not understood theoretically,
they may provide a useful testing ground for the deeper understanding of
dynamic nuclear polarization.

\end{abstract}

\vspace {2cm}


\vspace {2cm}

\newpage
\pagestyle{empty} 
\begin {Authlist}
B.~Adeva\Iref{a19},
E.~Arik\Iref{aa1}
S.~Ahmad\Iref{a17},
A.~Arvidson\Iref{a22},
B.~Badelek\IIref{a22}{a24},
M.K.~Ballintijn\Iref{a14},
G.~Bardin\Iref{a18},
G.~Baum\Iref{a1},
P.~Berglund\Iref{a7},
L.~Betev\Iref{a12}
% ...
\end {Authlist}
\Instfoot {a0} {The American University, Washington, D.C. 20016, USA}
\Instfoot {a1} {University of Bielefeld, Physics Department,
                33615 Bielefeld, Germany\Aref{i} }
\Instfoot {aa1} {Bogazi\,ci University and \,Ceknece Nuclear
                Research Center, Istanbul, Turkey }
\Instfoot {aa2} {Istanbul Technical University, Istanbul, Turkey }
\Instfoot {a2} {University of California, Department of Physics,
                Los Angeles, 90024~CA, USA\Aref{j}}
\Instfoot {a3} {University of California,
                Institute of Particle Physics,
                Santa Cruz, 95064 CA, USA}
\Instfoot {a4} {CERN, 1211 Geneva 23, Switzerland}
\Instfoot {a5} {University of Freiburg, Physics Department,
                79104 Freiburg, Germany\Aref{i}}
\Instfoot {a6} {GKSS, 21494 Geesthacht, Germany\Aref{i}}
\Instfoot {a7} {Helsinki University of Technology, Low Temperature
                Laboratory, 02150 Espoo, Finland}
\Instfoot {a8} {University of Houston, Department of Physics,
                Houston, 77204-5504 TX,
                and Institute for Beam Particle Dynamics,
                Houston, 77204-5506 TX, USA\AAref{j}{k}}
\Instfoot {a9} {JINR, Laboratory of Super High Energy Physics,
                Dubna, Russia}
\Instfoot {a10} {University of Mainz, Institute for Nuclear Physics,
                 55099 Mainz, Germany\Aref{i}}
\Instfoot {a11} {University of Mons, Faculty of Science,
                 7000 Mons, Belgium}
\Instfoot {a12} {University of Munich, Physics Department,
                 80799 Munich, Germany\Aref{i}}
\Instfoot {a13} {Nagoya University, Department of Physics, Furo-Cho,
                 Chikusa-Ku, 464 Nagoya, Japan\Aref{l}}
\Instfoot {a14} {NIKHEF, Delft University of Technology, FOM and Free
 University,
                 1009 AJ Amsterdam, The Netherlands\Aref{m}}
\Instfoot {a15} {Northeastern University, Department of Physics,
                 Boston, 02115 MA, USA\Aref{k}}
\Instfoot {a16} {Northwestern University, Department of Physics,
                 Evanston, 60208 IL, USA\AAref{j}{k}}
\Instfoot {a17} {Rice University, Bonner Laboratory,
                 Houston, 77251-1892 TX, USA\Aref{j}}
\Instfoot {a18} {DAPNIA, CEN Saclay, 91191 Gif-sur-Yvette, France}
\Instfoot {a19} {University of Santiago, Department of Particle Physics,
                 15706 Santiago de Compostela, Spain\Aref{n}}
\Instfoot {a20} {Tel Aviv University, School of Physics,
                 69978 Tel Aviv, Israel\Aref{o}}
\Instfoot {a21} {INFN Trieste and
                 University of Trieste, Department of Physics,
                 34127 Trieste, Italy}
\Instfoot {a22} {Uppsala University, Department of Radiation Sciences,
                 75121 Uppsala, Sweden}
\Instfoot {a23} {University of Virginia, Department of Physics,
                 Charlottesville, 22901 VA, USA\Aref{k}}
\Instfoot {a24} {Warsaw University and Soltan
                 Institute for Nuclear Studies,
                 00681 Warsaw, Poland\Aref{p}}
\Instfoot {a25} {Yale University, Department of Physics,
                 New Haven, 06511 CT, USA\Aref{j}}
\Anotfoot {a} {Now at CERN, 1211 Geneva 23, Switzerland}
\Anotfoot {b} {Now at University of Montreal, PQ, H3C 3J7,
               Montreal, Canada}
\Anotfoot {c} {Now at Max Planck Institute, Heidelberg, Germany}
\Anotfoot {d} {Permanent address: Miyazaki University,
               88921 Miyazaki-Shi, Japan}
\Anotfoot {e} {Permanent address: PSI, 8093 Villigen, Switzerland}
\Anotfoot {f} {Permanent address: KEK, 305 Ibaraki-Ken, Japan}

\Anotfoot {g} {Permanent address: University of Buenos Aires,
               Physics Department, 1428 Buenos Ai\-res, Argentina }
\Anotfoot {h} {Now at SSC Laboratory, Dallas, 75237 TX, USA}
\Anotfoot {i} {Supported by Bundesministerium f\"ur Forschung und
               Technologie}
\Anotfoot {j} {Supported by the Department of Energy}
\Anotfoot {k} {Supported by the National Science Foundation}
\Anotfoot {l} {Supported by Ishida Foundation, Mitsubishi Foundation and
Monbusho International Science Research Program}
\Anotfoot {m} {Supported by the National Science Foundation
               of the Netherlands}
\Anotfoot {n} {Supported by Comision Interministerial de Ciencia
               y Tecnologia}
\Anotfoot {o} {Supported by the US-Israel Binational Science
               Foundation and The Israeli Academy of Sciences}
\Anotfoot {p} {Supported by KBN}
\Anotfoot {q} {Now at Technische Universit\"{a}t, Dresden, D-01062}

\end {titlepage}

\newpage

	Measurements of deep inelastic scattering of polarized muons from polarized
protons and deuterons determine the spin dependent structure functions of the
nucleon which allow fundamental tests of quantum chromodynamics and of models
of nucleon structure \cite{hugh83}. The precision of these experiments is
strongly related to the polarization of the target nucleons. Therefore, the
large enhancement of our deuteron target polarization which we discovered
during the data-taking for deep inelastic muon scattering \cite{adev93} had a
significant impact on our experiment at CERN. The discovery was associated with
a faulty regulator of the high voltage power supply of a microwave source.
After controllable frequency modulation (FM) of the microwave tube was
implemented, a gain by a factor of 1.7 in the maximum deuteron vector
polarization and of 2.0 in the polarization growth rate were achieved. These
increases have been of crucial importance because data-taking extends over many
months and the statistical error is proportional to $1/PN^{1/2}$, in which
$P=<I_z>/I$  is the target vector polarization and N is the number of scattered
events. The magnitude of the enhancement has been reported earlier
\cite{adev94}. The purpose of this paper is to detail the full  characteristics
of this effect, to present new data on the electron paramagnetic resonance
absorption (EPR) spectrum and to discuss briefly processes which may contribute
to the FM phenomenon.

	The polarized target \cite{nini82,brown84} consists of two cells each $40~$cm
long and $5~$cm in diameter located in a large cylindrical multimode microwave
cavity. The two target halves are polarized in opposite directions by dynamic
nuclear polarization (DNP). The target material is glassy, perdeuterated
1-butanol, $\rm{C_4D_9OD}$ with $5\%$ by weight of deuterium oxide, doped with
the paramagnetic EDBA-Cr(V) complex \cite{krum90} to a concentration of $7\cdot
10^{19}~cm^{-3}$ \cite{bult94}. It is located in a magnetic field of $2.5~$T
with a uniformity of $10^{-4}$ over the volume and is cooled by a dilution
refrigerator. The DNP is obtained by applying microwave power near the EPR
frequency of the paramagnetic complex.

	The deuteron vector polarization is measured with nuclear magnetic resonance
(NMR) probes, each of which is part of a series tuned Q-meter circuit
\cite{gcou94}. The material is sampled by five probes in each target cell. The
polarization is determined from the integrated NMR signals, calibrated in
thermal equilibrium at 1K. The relative accuracy of the measurement $\delta
P/P$ is $5\%$ \cite{adev94}.

	The microwave power for DNP is produced by two extended interaction
oscillators (EIO) with an emission bandwidth of about $0.1$~MHz. The rate of
polarization is optimized by controlling the microwave power and frequency. The
frequency is controlled by the EIO cathode voltage with a sensitivity of about
$0.4$~MHz/V or by tuning the EIO cavity. The power is controlled by
non-reflective attenuators.

	For materials in which the solid effect \cite{abra58,schm65} dominates as a
mechanism for DNP, it has been found that microwave FM can improve the rate of
DNP. This appears to result from the fact that FM counteracts the effect of
``hole burning'' due to EPR absorption at a fixed frequency \cite{kest85}. In
the glassy alcohol materials with Cr(V) complexes, where the dynamic nuclear
cooling \cite{abra64} is the dominant mechanism for DNP, hole burning is not
expected. However a polarization enhancement of $10\%$ to $20\%$ was observed
in a fluorinated alcohol leading to polarizations of $\approx 0.80$ for protons
and $^{19}$F \cite{dahi89}. References to enhancements of a few percent at
polarizations around $0.70$ or of about $15\%$ for a material with only a few
percent polarization can also be found in \cite{dahi89}. To our knowledge no
studies have been reported for the effect of FM on deuteron polarization except
in one case where FM was used to compensate for magnetic field inhomogeneity
and improve the final polarization by $5\%$ to $\approx 0.30$ \cite{ishi89}.

	The large enhancement of deuteron polarization in our target due to FM came
therefore as a surprise. Figure~1 shows the typical time evolution of the
deuteron polarization $P_D$ without and with FM. For this figure the cathode
voltages were modulated at $1~$kHz with a $\approx 50~$volt peak-to-peak
amplitude leading to a FM amplitude $\Delta f \approx 20~$MHz  for the $69~$GHz
microwave source. The maximum deuteron vector polarizations under these
conditions were $0.43$ and $-0.49$.

	The EPR spectrum was measured in our target at a constant frequency by
scanning the magnetic field. Such a spectrum, shown in Figure~2 without FM, was
obtained using a $220~\Omega$ Speer composite carbon resistor as a bolometer,
located in the dilute phase of the mixing chamber outside the target material
\cite{niin79}. The input power to the microwave cavity $\dot{Q}_{IN}$ is the
sum of $\dot{Q}_{MAT}$, the power absorbed by the material in the EPR process,
and $\dot{Q}_{NR}$ the non-resonant power absorbed into the cavity. The power
absorbed by the bolometer $\dot{Q}_{SP}$ is a constant fraction $r$ of
$\dot{Q}_{NR}$. It can be expressed as $\dot{Q}_{SP}=c(T^4_{SP} - T^4_{HE})$
where $T_{SP}$ is the temperature of the bolometer, $T_{HE}$ is the temperature
of the dilute phase and $c$ is a constant \cite{loun74}. During the EPR
measurement the input power $\dot{Q}_{IN}$ remains constant and we can neglect
the variations of $T^4_{HE}$. Consequently the relation $\dot{Q}_{IN} =
\dot{Q}_{MAT} + \dot{Q}_{NR} = \dot{Q}_{MAT} + (c/r)\times(T^4_{SP} -
T^4_{HE})$ shows that $\dot{Q}_{MAT}$ is a linear function of $T^4_{SP}$. The
broad absorption band seen in Figure~2 is due to the anisotropy of the
$g-$factor of the EDBA-Cr(V) electron spin. The highest positive and negative
polarizations without FM were obtained at frequencies $f^+_0=69.090 ~$GHz and
$f^-_0=69.520 ~$GHz, respectively.

	The EPR spectra with better resolution at the edges of the absorption band are
shown in Figure~3a and 3b, both with and without FM. The data points with FM
were obtained using a modulation amplitude $\Delta f=4~$MHz to keep a good
resolution in our spectra. In order to measure the small change in EPR
aborption due to FM a novel technique of making consecutive measurements of the
bolometer resistance with and without FM at each field step was employed. In
Figures~3c and~3d we display the difference $\Delta T^4_{SP} = (T^{{\sf
off}}_{SP})^4 -  (T^{{\sf on}}_{SP})^4 = ( \dot{Q}^{\sf on}_{MAT} -
\dot{Q}^{\sf off}_{MAT})/{\it c}$ . These data demonstrate that FM increases
$\dot{Q}_{MAT}$ in the edges of the EPR spectrum. Note that the structures in
Figure~3a and 3b which extend down to $69.00~$GHz and up to $69.60~$GHz are
almost entirely eliminated in the presence of FM even though the amplitude of
FM is small compared to their width.

	In Figure~4 we show the difference $\Delta T^4_{SP}$ as a function of the
frequency of FM for different input power levels $\dot{Q}_{IN}$ with an FM
amplitude $\Delta f = 30~$MHz at $69.090~$GHz where $\Delta T^4_{SP}$ reaches a
maximum. This difference grows with the modulation frequency up to a maximum
value (indicated by the arrows) and then remains constant. The frequencies at
which the additional EPR absorption reaches its maximum value increase roughly
linearly with $\dot{Q}_{IN}$. A study of the polarization growth rate was
performed at high negative $P_D$ values for a setting of $\dot{Q}_{IN}$ close
to the one which was used for curve 2 of Fig.4. The rate increased with
modulation frequency and reached a maximum value of $-0.8\%$ per hour when
modulating at $10~$Hz. At this $\dot{Q}_{IN}$ value, $\Delta T^4_{SP}$ reaches
a maximum at this frequency which suggests strongly that the additional EPR
absorption due to FM is what leads to the enhanced DNP.

	In further measurements, we have established that the highest positive and
negative polarizations with FM were obtained using $\Delta f\approx30~$MHz at
$f^+_0=69.070~$GHz and $f^-_0=69.540~$GHz, respectively. The gain in maximum
polarization due to FM is 1.7 and the increase in polarizing speed is about
two. The homogeneity of the deuteron polarization throughout the target volume
was investigated . Two radially superimposed coils measuring polarization at
radii of $1.5~$cm and $0.5~$cm showed a deuteron polarization ratio from the
small to the large coil of 1.20 before and 1.06 after applying FM
\cite{adev94}. A study of the deuteron NMR line asymmetry \cite{adev94}
provided us with an upper limit $\Delta P_D$ for the spatial variation of
$P_D$. Typical values for $\Delta P_D$ were $0.30$ without FM and $0.15$ with
FM which confirmed that FM improves the uniformity of polarization.

	The large enhancement has been confirmed recently in our new $2.5~l$ target
\cite{kyyn94} where maximum deuteron polarizations of 0.46 and -0.60 were
observed using FM. In spite of the improved magnetic field uniformity of
$3.10^{-5}$ of the new target, the optimum $\Delta f/f$ for FM was about
$0.5.10^{-3}$ as in the previous target with  $10^{-4}$ field uniformity. Also
we find that $\Delta f/f$ is large compared to the target magnetic field
inhomogeneity in both targets. We conclude that the mechanism by which FM
improves polarization has little to do with the field nonuniformity. For
protons FM increased the polarization, typically from 0.75 to 0.85, and to
maximum values as high as 0.95 \cite{adam94}.

	The existing theory \cite{prov62a,prov62b,kozn64} provides a qualitative
understanding of the DNP for our target material; however, the large
polarization enhancement due to microwave FM may require additional mechanisms.
An example is the cross-relaxation within the system of electron spins which
has been assumed to be fast. It has been suggested \cite{niin92} that a slow
cross-relaxation may lead to a lack of thermal equilibrium among electron
spins and hence to unequal spin temperatures for different nuclei which results
in lower nuclear polarization. FM may counteract this effect by increasing the
number of electron spins which are saturated.

	A possibly related effect is the local depletion of the electron spin packets
which has been observed for materials whose EPR lines are broadened by
hyperfine interactions when irradiated at fixed frequency. With FM, this local
depletion can be avoided and a migration of spin packets occurs towards the
wings of the EPR band \cite{gold69}. This may result in a stronger EPR
absorption in the wings.

	Since the aim of our experiment was to measure spin-dependent asymmetries in
polarized deep inelastic scattering, we did not attempt a more detailed study
of the effect of FM on the target polarization. Our observations of the EPR
absorption were used as a guide to optimize the parameters of the FM.

	In conclusion, we discovered a large increase in deuteron polarization due to
frequency modulation which is of great value for our high energy physics
experiment. We found that an amplitude of FM of $\approx 30~$MHz and a
frequency of $1~$kHz improved the deuteron polarization growth rate by a factor
of 2 and resulted in deuteron polarizations as high as $0.60$ with improved
spatial uniformity over the target volume. Relations of this large FM effect to
features of the EPR absorption mechanism were found and may provide useful
information to a deeper understanding of dynamic nuclear polarization.

\newpage

\vspace{0.5cm}

{\bf Figure Caption}

\vspace{0.5cm}

{\bf Figure 1}

\vspace{0.5cm}

Deuteron polarization as a function of time without FM (dark circles) and with
FM (open circles). Positive and negative polarizations are shown.

\vspace{0.5cm}

{\bf Figure 2}

\vspace{0.5cm}

Electron paramagnetic resonance absorption band for the glassy perdeuterated
butanol of the SMC polarized target (the dotted line guides the eye). The
temperature $T_{SP}$ is derived from the value of a Speer carbon composite
resistor located near the material. The measurements were performed at a
constant frequency $f_0=69.520~$GHz by stepping the magnetic field. The field
values $H$ are converted to the equivalent frequencies $f=f_0H/H_0$ at
$H_0=2.5~T$.

\vspace{0.5cm}

{\bf Figure 3}

\vspace{0.5cm}

Enhancement in the wings of the EPR absorption spectrum observed when the
microwave frequency was  modulated with an amplitude of $4~$MHz at $1~$kHz
frequency. Figures a and b show the EPR spectra obtained without (dark circles)
and with FM (open circles) for the domain of frequency leading to positive (a)
and negative (b) polarizations. Figures c and d show the differential effect.

\vspace{0.5cm}

{\bf Figure 4}

\vspace{0.5cm}

Enhancement of the EPR absorption as a function of the FM frequency for
different values of the input microwave power ${\dot{Q}_{IN}}$. The arrows show
the frequencies at which the maximum enhancement is reached. The four curves
labelled 1, 2, 3 and 4 were obtained at levels of input power $\dot{Q}_{IN}$
increased successively by a factor 4.
\newpage
\begin{center}
{\bf FIGURE 1}
\end{center}
\vspace{3cm}
\begin{figure}[here]
\epsfig{figure=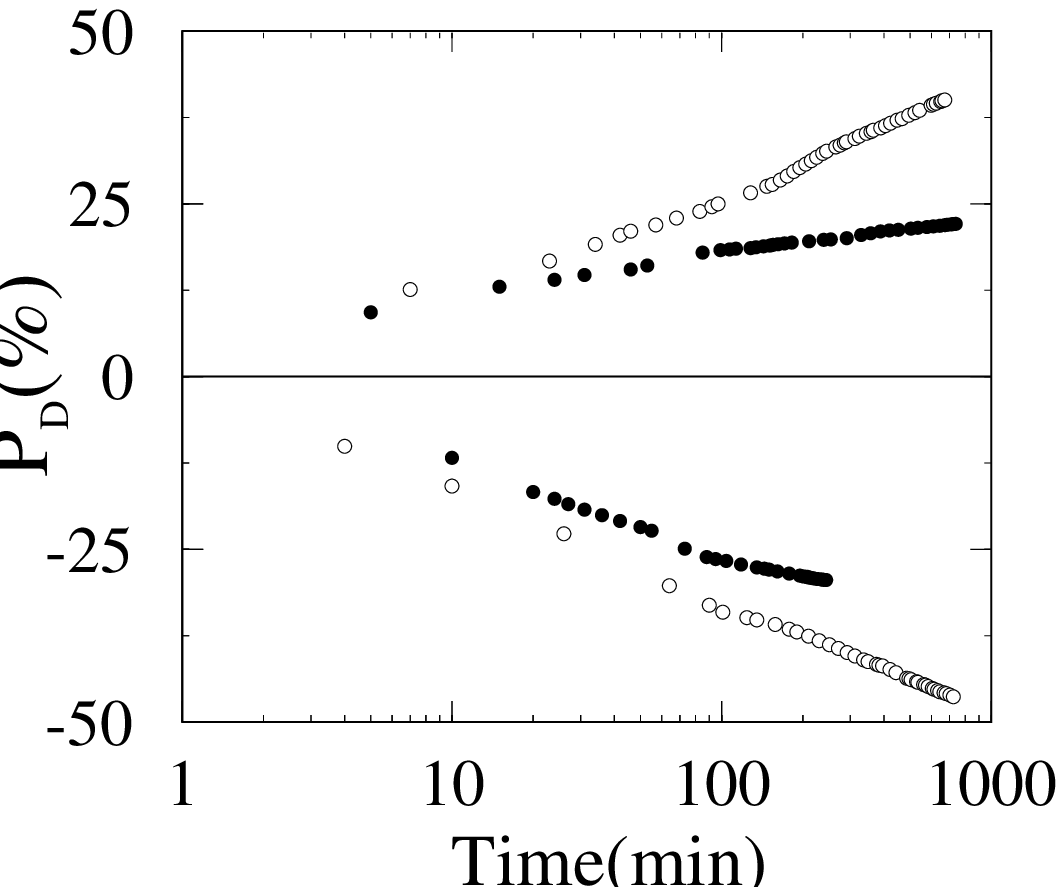,width=1.3\textwidth}
\end{figure}
\newpage
\begin{center}
{\bf FIGURE 2}
\end{center}
\vspace{3cm}
\begin{figure}[here]
\epsfig{figure=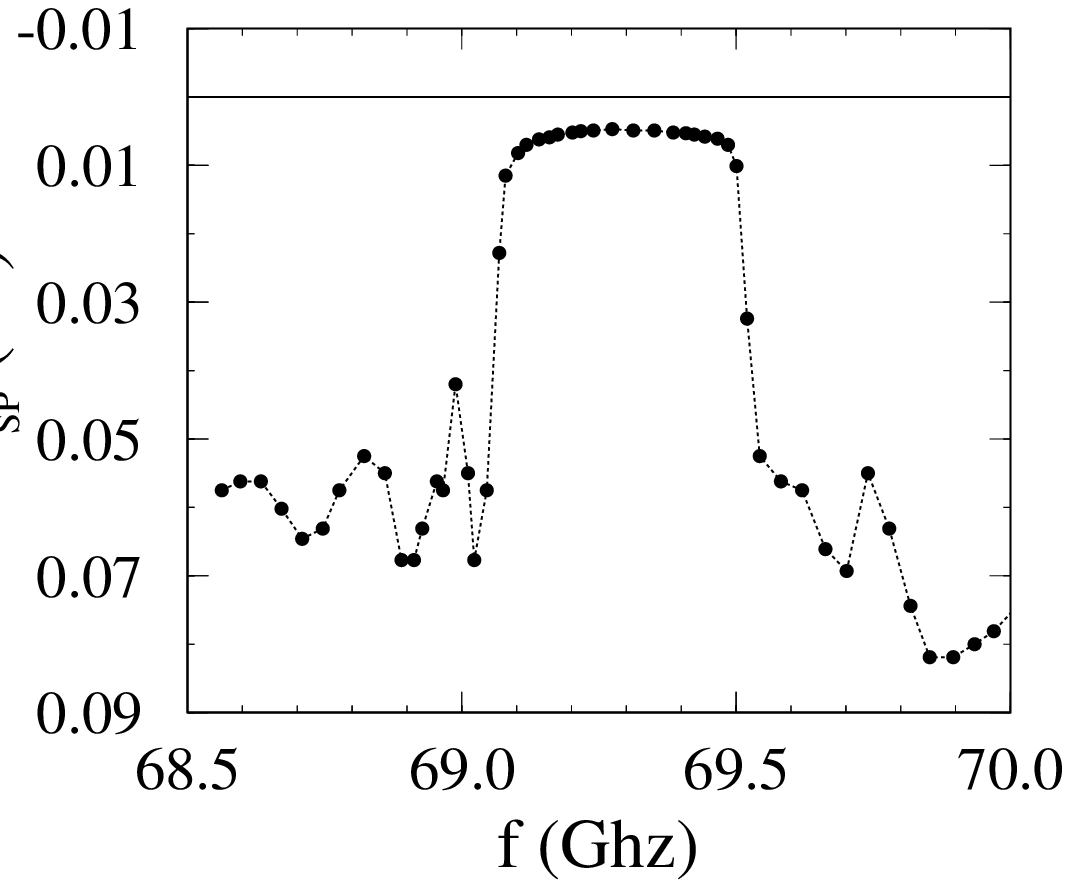,width=1.3\textwidth}
\end{figure}
\newpage
\begin{figure}[t]
\begin{center}
\begin{center}
{\bf FIGURE 3}
\end{center}
\vspace{1cm}
 \mbox{\begin{tabular}[t]{cc}
    \hspace{-2.5cm}
     \subfigure[ ]%
		{\epsfig{figure=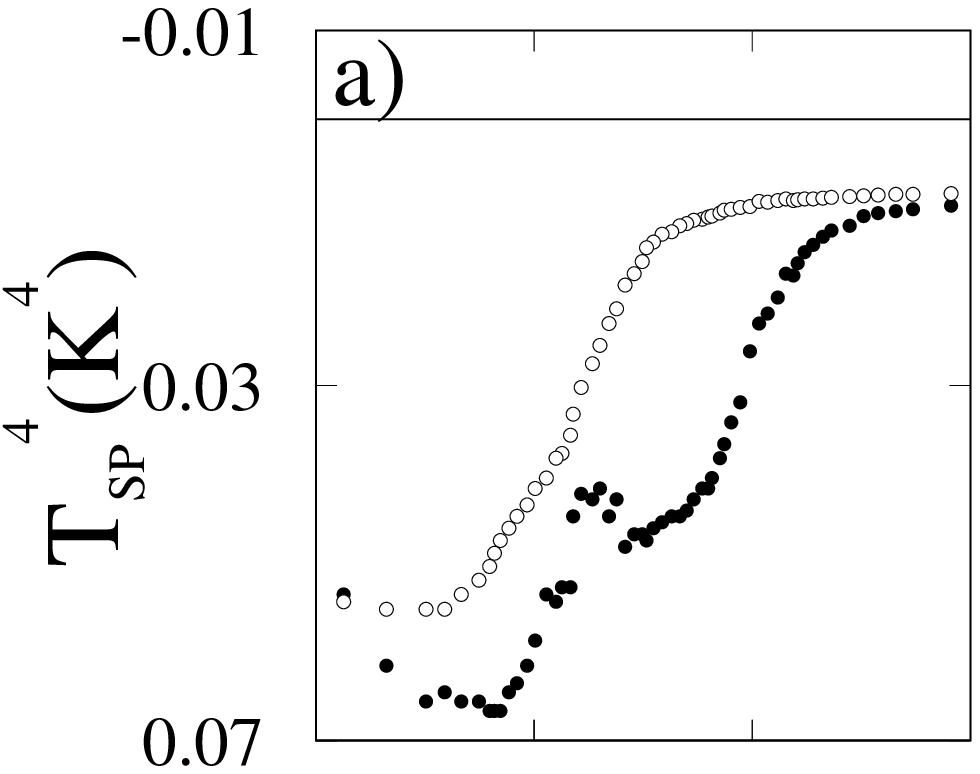,width=0.9\textwidth}} &
    	\vspace{-1cm}
	\hspace{-6.cm}
     \subfigure[ ]%
		{\epsfig{figure=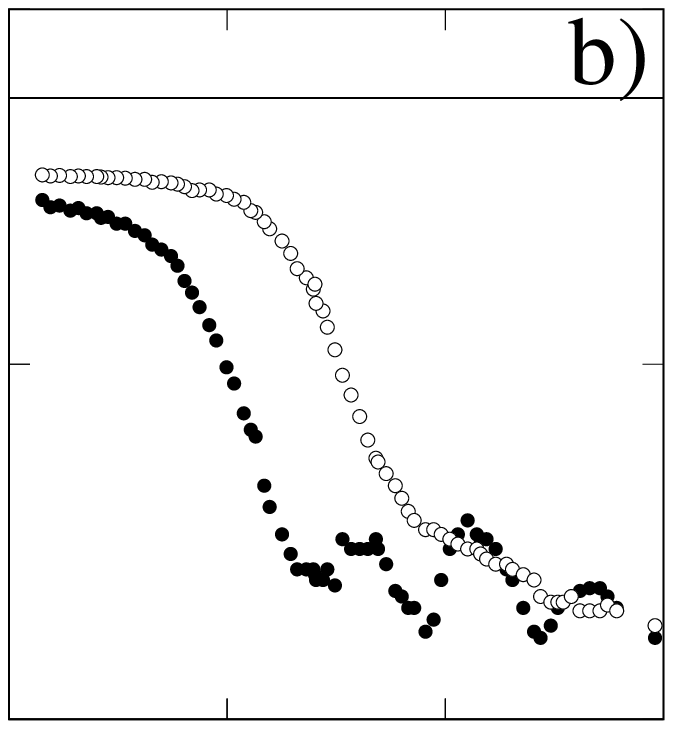,width=0.9\textwidth}} \\
    \vspace{0cm}
    \hspace{-2.5cm}
     \subfigure[ ]%
		{\epsfig{figure=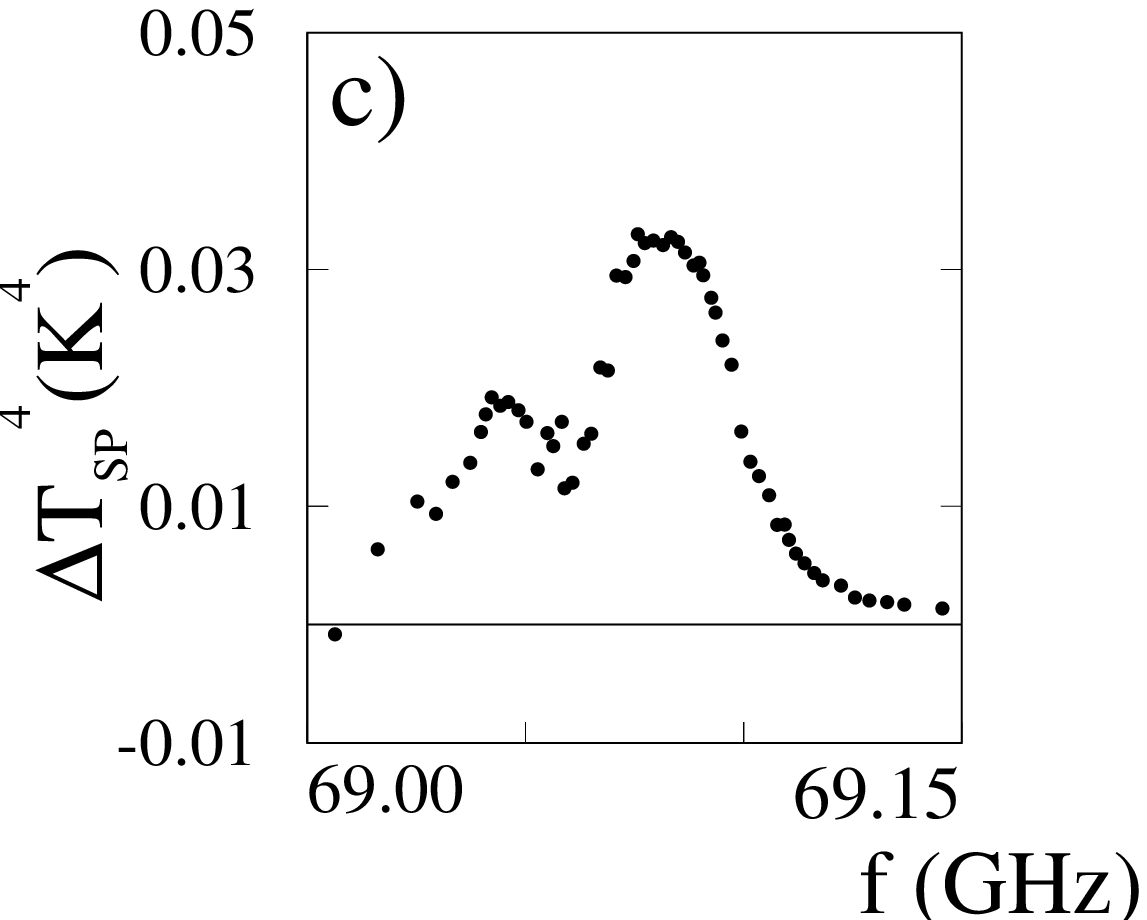,width=0.9\textwidth}} &
    \hspace{-6.cm}
     \subfigure[ ]%
		{\epsfig{figure=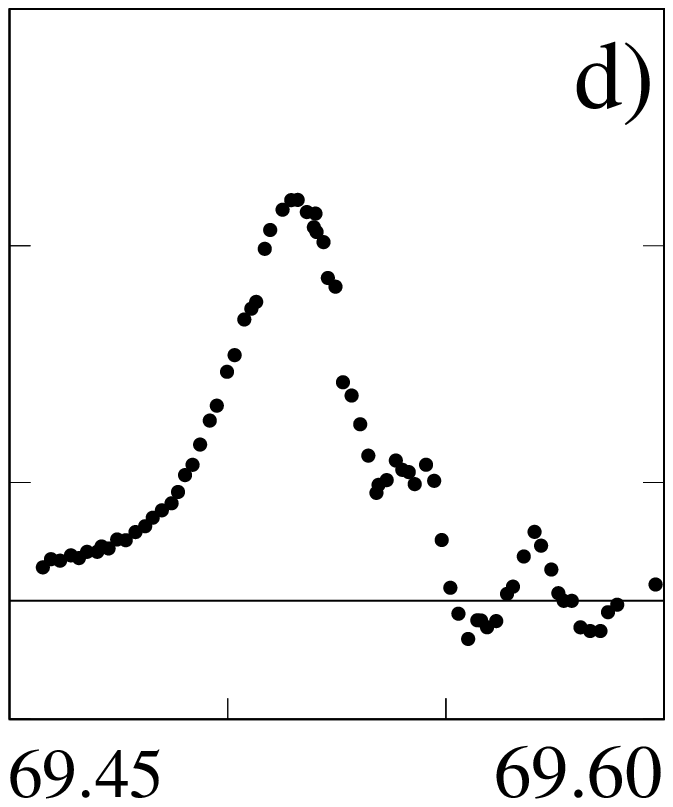,width=0.9\textwidth}}
\end{tabular}}
\end{center}
\end{figure}
\newpage
\begin{figure}[here]
\begin{center}
{\bf FIGURE 4}
\end{center}
\vspace{3cm}
\epsfig{figure=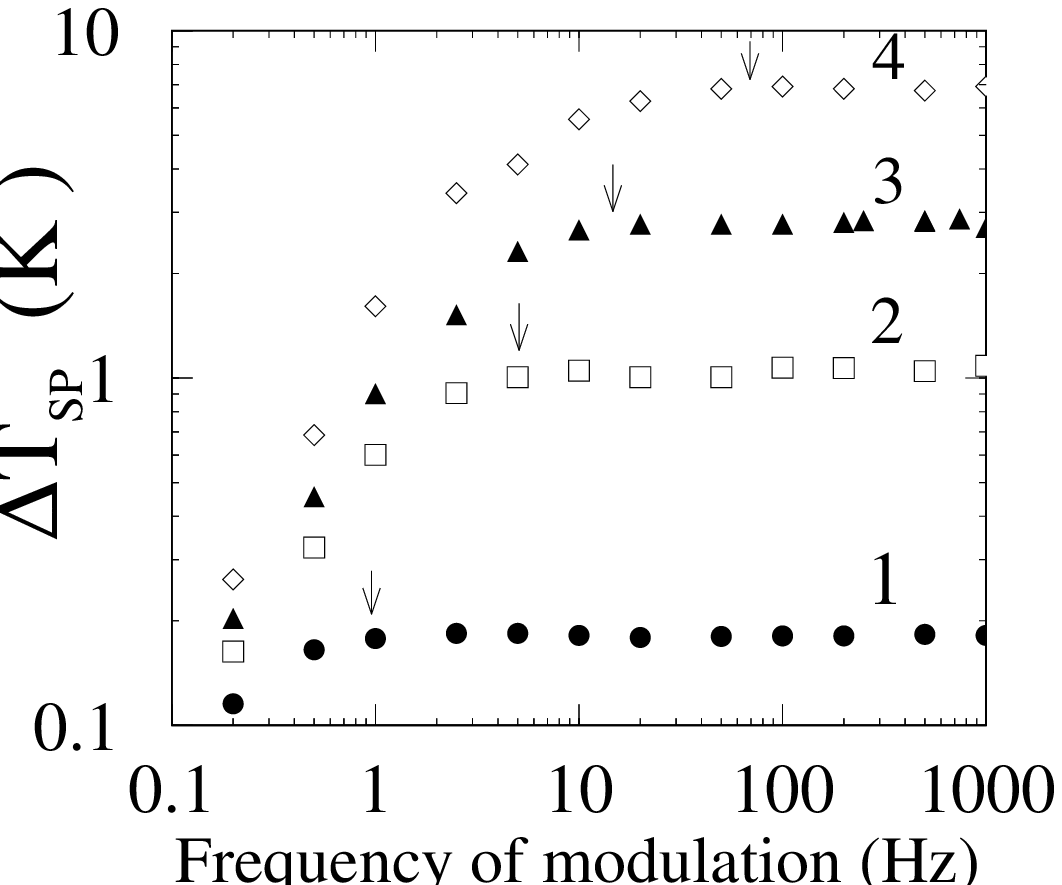,width=1.3\textwidth}
\end{figure}
\end {document}